Agent-Based Models in Social Physics

Le Anh Quang, Nam Jung, Eun Sung Cho, Jae Hwan Choi, and Jae Woo Lee*

Department of Physics, Inha University, Incheon, Korea

Abstract

We review the agent-based models (ABM) on social physics including econophysics. The ABM consists of agent, system space, and external environment. The agent is autonomous and decides his/her behavior by interacting with the neighbors or the external environment with the rules of behavior. Agents are irrational because they have only limited information when they make decisions. They adapt using learning from past memories. Agents have various attributes and are heterogeneous. ABM is a non-equilibrium complex system that exhibits various emergence phenomena. The social complexity ABM describes human behavioral characteristics. In ABMs of econophysics, we introduce the Sugarscape model and the artificial market models. We review minority games and majority games in ABMs of game theory. Social flow ABM introduces crowding, evacuation, traffic congestion, and pedestrian dynamics. We also review ABM for opinion dynamics and voter model. We discuss features and advantages and disadvantages of Netlogo, Repast, Swarm, and Mason, which are representative platforms for implementing ABM.

1. Introduction

Since a society composed of many people is a typical system of many-body, it is possible to apply the principles of statistical physics and complex systems. Recently, there has been a growing interest in social physics to understand social phenomena from the viewpoint of complexity [1-10]. Society represents diverse structures, patterns, and organizations in spite of the absence of universal designers to design the whole. The interaction between people who make up society has resulted in various organizations such as schools, businesses, countries, etc. Social physics is a new discipline that understands emerging phenomena in society by applying theories and methods of statistical physics and complex science to social phenomena based on interaction among people.

The agent-based model (ABM) is a very useful research method when studying social systems [3-10]. The agent-based model refers to a system of many-body that exhibits emergent characteristics when autonomous agents interact with one another. Each agent is a group of people with a set of rules that make autonomous decisions and interact with each other. Agents also interact with the environment surrounding them. Agents interact with their environment and neighbors according to a given set of behavior rules, and then decide how to act [3-4]. If the agent is an economic entity participating in the market, then he or she can produce, consume, or sell any things when they participate to the market [1-25]. Since repetitive interactions among many agents can be easily implemented by computer simulations, agent-based models evolve with the development of computers.

The ABM is a model that reflects the interaction of agents, thus enabling a natural description of the system. Complex systems can not represent the whole system as a sum of parts. ABM can reproduce the emergent characteristics of this complex system by computer simulation [1-10]. For example, traffic flow on the highway is affected by each driver's individual driving habits, but phenomena such as phantom traffic jams on the highway are emerging patterns as a complex feature of the interacting drivers [5-10]. The ABM is also a flexible model. Models can easily be improved by changing the attributes of agents or by adding features. The ABM describes the system in a bottom-up rather than a top-down manner. Because ABM is diverse, one can classify them following by various methods and aspects. According to the characteristics of the system, ABM can be classified into the four categories such as flow models, market models, models of organizations, and diffusion models [1-4]. The flow model relates to the flow of agents in a social or economic system. Typical examples include traffic flow, evacuation from buildings, customer flow management, and so on. Market models include socio-economic markets or organizations. Stock market, strategic simulation, system risk or operational risk are typical starting models. The models of organizations are related to spontaneous appearances of structure or organization in the society. Diffusion models consider diffusional processes in society. The spread of innovation, the spread of advanced science and technology, and the spread of computer viruses online are among the diffusion models [1-10].

In this article we review the current progress of the agent-based model on social physics. We discuss the basic concepts of the agent-based model in Section 2. The ABM is consisting in agents, system space, and external environments. In Section 3 we review the current works of the ABM on the social complex systems. We introduce the platforms of ABM. In Section 4 we gave the concluding remarks.

2. Agent-Based Model

The agent-based model consists of three elements: the *agent*, the *system space* in which agents move, and the *external environment*. Agents influenced by the external environment behave autonomously and interact with each other in the system space. The agent-based model is introduced to explain the socio-economic phenomenon in the early days and the application has been expanded with the development of computers. Many applications of the AMBs are conducted in the fields of agent-based computational economics applying agent-based models to economic phenomena, agent-based computational sociology applied to sociology, and individual-based ecology applied to ecology [1-5, 11].

The basic components of the agent-based model are as follows.

(1) Agent: Agent has attributes and autonomous rules of behaviors.

(2) System Space: Agents are defined on a space and they interact with other agent and environment in this space.

(3) External environment: Agents are surrounded by the external environment and they are influenced from their environment.

2.1 Agents

Agents are the basic building blocks of systems in ABM. Each agent is identifiable and acts autonomously according to established rules. In the study of social phenomena, the people who constitute society become agents, and in the researches dealing with multi-particle systems, they become the constituents of the system. The most basic feature of an agent is to act autonomously. In other words, it acts autonomously according to its own rules given without any external instructions on a given situation. The characteristics of agents in ABM are summarized as follows [1-11].

- *Autonomous*. Agents behave autonomously. Autonomy means that an agent does not take central control and act in his own judgment. The agent makes an independent decision by synthesizing his/her own information, information obtained from interaction with neighbor agents, and information obtained from external environment. There is no a supervisor, a conductor, or a dictator who commands the entire system[1-4]. In the movement of birds, there is no a controller, and each bird behaves according to his own independent judgment. Of course, when agents interact with neighbors' agents, there may be limitations such as recognizing only a limited distance or not knowing all the information of the system. Agents therefore act actively, not passively.

- *Interaction.* Agents interact. The agent interacts with other agents, and environment. In the model, the agent performs a local or global interaction with neighboring neighbors in the system space. A company employee who comes to the company decides the behavior by interacting with the people of the business colleagues, superiors, and customers in the vicinity. In current world, where social networks are actively used, users who use SNS or Twitter express their opinions by actively interacting with their neighbors on the network connected to them.

- *Rules of Behaviors.* Agents act in accordance with the rules of behaviors. Agents perform nonlinear interactions and, in some cases, have thresholds when they act. In the agent-based model, each agent has activity according to its own action rule and its interaction with neighboring neighbors. This agent is called a minimalistic agent if the agent assumes no intelligence [4,26]. Minimal agent models are often found in models with minimal interaction between agents in the field of physics or natural sciences. In the opinion dynamics model, a person with a positive opinion is expressed as S = + 1 or up spin, and an agent with a negative opinion is expressed as S = -1 or down spin [1-4,27-31]. The spin opinion dynamics model is a representative minimal agent model. On the other hand, agents have complex agents with proactivity and reactivity [4]. There is also a multi-agent system, in which agents with totally different attributes are gathered.

- *Irrationality.* Agents often act irrationally. If the agent is a human, there are situations in which he acts irrationally. It is natural for agents to act with bounded rationality because they do not know the information of the system as a whole. The concept of rationality is that all agents can know all the information of the system and have complete interpreting ability so that they can make the most reasonable choice for themselves. In economics, a rational agent has all the market information about the object when he/she buys it, so that he can know where he/she is most likely to benefit from the price of the object, and purchases the object at the shop [28-31]. A rational agent is an optimizer with infinite capabilities. Many economists assume the rational market composed of such rational agents and establish theories [28-32]. In daily life, however, agents do not know all the information about the system. Agents have limited rationality to act with limited ability only with limited information they can obtain.

- *Memory and learning.* Agents have memories of the past and behavior of agents depend on the historical path [1-5]. Non-Markovian behaviors show behavior with temporal correlation. Agents also learn through experience. Agents constantly adapt to the surrounding environment. Memory and learning are important for agents when they adapt to the environment. A new employee adapts to the company quickly. When an ecological change occurs, the creature adapts to the environment. Adaptation occurs at the agent's personal level. They may not fully adapt to the changed environment. Individuals may have different adaptive abilities, so adaptation speeds vary with individual differences. Adaptation can also occur in the adaptation of population or social units.

- *Heterogeneity*. Agents are heterogenous. Each agent has his own attributes and rules of behaviors. In the social network, each agent has various attributes to express his/her opinion. There are progressive people and conservative people. In birds' flock, birds have their own position and orientation. It is common for agents in the agent-based model to have heterogeneity rather than homogeneity. The agent-based model is a bottom-up model because the attributes or behavior rules of individual agents indicate the emergence of the system [1-10].

2.2 System space

Each agent can interact with other agents. They interact with each other in the system space. If each agent has a physical position, it moves and affects each other in a given space. If the agent is on a specific network, the agents interact with each other according to the network connection status. The space in which the agents are located is called the system space. The system space is either a physical space that simulates the real world or an abstract space such as a network [1-16]. Many agent-based models are implemented in a two - dimensional regular lattice. However, the spaces in which agents move in social phenomena have various structures. In the case of a person moving in a city, it moves in a continuous space rather than a lattice structure. Recently, the position of a person can be expressed in a real space by using a GPS (Global Positioning System). The movement of a person in the city can be displayed on the city map using the location information of the mobile phone. Traffic flows and escape from buildings map on regular structures such as roads and corridors [1]. In the formation of an opinion or an election model, the system space considers a square lattice or network structure [3,4]. Interaction with the nearest neighbors is important in ABM. Use the von Neumann neighborhood or the Moore neighborhood to determine neighboring neighbors as shown in Fig. 1. The boundaries of the network usually use periodic boundary conditions or closed boundary conditions. A network such as the Internet provides another connection space from the grid space or geographic information space.

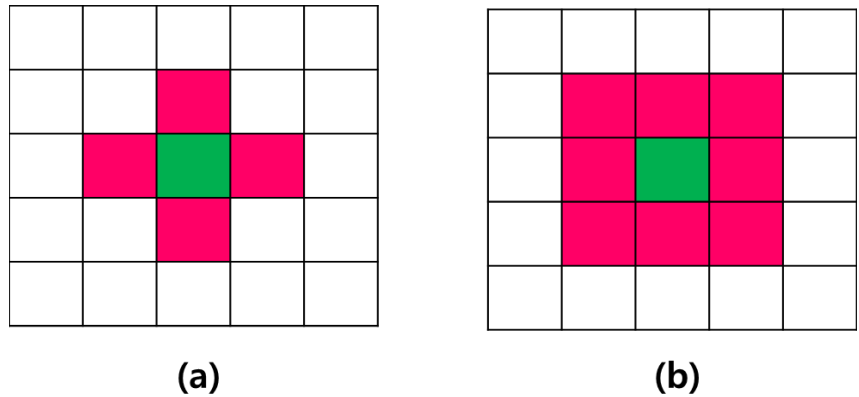

Fig. 1. We use boundary conditions as (a) von Neumann neighborhood or (b) the Moore neighborhood in the ABM simulation.

A network such as the Internet or a power grid network is formed as a physical network, but a user using the social media use the network without knowing the conneciton of the physical network. A network can be classified into a regular network or a lattice, a random network (RN), a small-world network (SWN), a scale-free network (SFN), a hierarchical network (HN), a modular network (MN) [33-40]. In recent years, research on a bipartite network, a multiplex network, and a multilayer network is leading interests as well. A network excluding a regular network is called a complex network [33,34]. We simulate the AMB on many types of complex networks according to the models.

2.3 External environment

Agents are affected by external agents as well as other agents. In social phenomena, agents influencing members such as the press, social media, stock indices, and exchange rate of financial market etc. can be considered [1-10]. In the case of natural phenomena, external agents such as air pressure, humidity, and temperature influence to the agents in the system. In the agent-based model, agents are the constituent units that make up the system, and they are different for each system. Examples of agents include people, animals, cells, automobiles, and economic entities etc [1,2]. These agents are not only influenced by the interactions among the agents within the system, but also dynamically change their behavior under the influence of the external environment [4]. The effect of the environment on the system can be divided into two major ways. The first is that external influences are local to some part of the system. An example of *local* external influences is rumor propagation. The rumors that originated in a single source of cheating spread throughout the system as they spread to the neighbors. The more people who have a tendency to spread rumors in society, the easier it will spread to the whole society. The second is that external influences act *globally* throughout the system. An example of the global external influence is the case where information is transmitted across the stock market or the foreign exchange market. If the US Federal Reserve decides to raise interest rates today, the news will be broadcast worldwide. Tomorrow, investors who participate in the stock market will know that information and they buy or sell stocks.

3. Agent-Based Models on Social Complex Systems

3.1 ABM for Human Behaviors

Social physics is an area that explores the complexity of social systems. Although it can include economic physics in a broad sense, this article looks at the latest research trends in the narrow sense of social physics [1,2,5-10]. Bonabeau classifies agent-based model simulations as flow simulation, organizational simulation, market simulation, and diffusion simulation [4]. In 1971, Schelling proposed a segregation model to explain the natural separation of races in the city [41]. In this model, agents are satisfied if their neighbors are equal to t%, and leave the area. When t increases in the two-dimensional lattice space, isolation between agents occurs naturally. This isolation model explains the isolation of races in real cities. In 1996, Epstein and Axtell proposed a civil violence model that simulates violence in society [42,43]. It opens up the possibility of explaining decentralized rebellion and inter-ethnic civil violence with minimal control variables. Axelrod explained the emergence of cooperation in society based on the prisoner's dilemma 'game theory [44]. These agent-based models help to understand individual behavior in society. It is possible to explain why individuals' choices in society are deviated from the rational choice theory [45]. It also opens the possibility of explaining social co-operation such as beliefs, reciprocity, and reputation among people in society and the emergence of social order in a simple model [46]. While some communities, organizations and relationships in society have cooperation and solidarity, some societies can understand the phenomenon of antagonism and division as an agent-based model [47].

The emotion contagion is one of the interesting issues in the online social media and real life. Several agent-based models for the emotion contagion have been proposed to explain the cooperation on the complex networks [48,49]. Reputation influences on the formation of opinion. There are controversies for formation of cooperation by the reputation of the agents [50-52].

3.2 ABM in Econophysics

In 1996, Epstein and Axtell proposed a virtual world agent-based model in which agents occupy resources (virtual sugar) in a limited space and adapt to a limited environment. This model is known as the Sugarscape model [52,53]. In this model, agents develop their wealth (the amount of sugar) starting from the very first environment. The Sugarscape model is a very simple model, but computer simulations show a imbalance of wealth [54]. The Sugarscape model consists of a 51x51 cell grid with sugar mountain located at two diagonal points. At each time step, the agent searches for his neighbor cell and then moves to the place where the sugar is most and takes sugar. When the sugar in the cell is depleted, it returns to the maximum value according to the recovery rate. Each agent has a movement rule and a metabolic rate, so he needs a minimum amount of sugar to survive. The agent below the metabolic rate dies and lays a child on the spot. The new born agent inherits parental information, but the metabolic rate and initial sugar content are arbitrary. The asymmetry of wealth can be found in the sugarscape model [54]. At the initial stage of the simulation, the agent's wealth is equal. In the early stage of the simulation, the distribution of wealth has a small number of poor agents and rich agents, and a wide middle-layer distribution. As the simulation progresses, a small number of agents near the peaks of Sugar Mountain begin to take up a lot of sugar. Agents located on the outskirts of the Sugar Mountain have only enough sugar to bear life. Agents are increasingly gathering near the peaks of Sugar Mountain and appear agents with a lot of wealth. The wealth asymmetry occurs by itself.

In many countries, the wealth distribution function owned by each individual shows a pattern. The distribution of wealth owned by an individual follows the 'Pareto distribution' [8-24]. Pareto's law is an empirical rule in which the distribution of wealth decreases along with the power law. A feature of the power law is that the number of people with large wealth is considerably greater than the normal distribution. In addition, the proportion of wealth that these minorities occupy is a large part of the total wealth. As is commonly known by the 80-20 rule, 20% of the rich people own 80% of the total wealth [8-24]. In the Sugarscape model, the distribution of wealth does not follow the perfect Pareto law, but the phenomenon of wealth inequality is observed and the capitalist economic system emerges spontaneously. In the Sugarscape model, wealth inequality does not depend on the details of the model. For example, wealth inequality is still observed even if the metabolic rate of the agents is changed or the simulation is carried out by varying the distribution of the sugar initially given to the agents. It is a characteristic of the model itself and wealth inequality is the result of interaction among agents. That is, it is a unique characteristic expressed in the system. The original Sugarscape model was transformed into a model with sugar and spice, and expanded to a model with added market function.

Holland and Miller proposed artificial adaptive agents in model of the economic market [40]. The Santa Fe Institute introduced an agent-based model of the stock market which is known as the

Santa Fe Artificial Market Simulator [41]. Lux and Marchesi observed scaling and criticality in a stochastic multi-agent model of a stock market [42].

The agent-based model of the financial market suggested some possibilities of explanation for the main characteristics of the economic system. The agents are acting rational or irrational. The efficient market hypothesis is based on the concepts of the rational agents and equilibrium market. The agent-based model can reproduce some properties of the financial market such as the stylized facts of the market. We can validate the results of the ABM by the real data such as the financial time series of the stock market, the prices of the commodities, foreign exchange rate, etc.[8,9,14,43-47].

3.3 ABM in Game Theory

Another successful model of the agent-based model is proposed by Challet and Zhang in 1997 [48]. Minority game model is a form of evolution game. The game consists of N (odd) agents. Each time step, each agent chooses A or B ($S_i = +1$ or $S_i = -1$). It is a game in which all agents win one of the independents and then the minority win. The simplest game is that the agents who won the minority title get one point. There are various ways of giving a score. How will each agent make his decision? Each agent makes a judgment based on past records. I think the past record simply records the winner. Therefore, the winning side can be represented by 1 or 0. That is, 1 indicates that A has won, and 0 indicates that B has won. The decision strategy of minority game is as follows. Each agent has limited ability when participating in the game. Assume that agents memorize only the most recent M-bit information when participating in a game win or loss. For example, if M = 3 bits of memory, the number of possible win / loss is $2^M = 8$. Agents use an array of possible win and lose numbers to record their strategy of choosing A or B in the next game. Thus, the total number of strategies that an agent can have. As the number of memory bits increases, the number of possible strategies increases exponentially. The agent takes only S strategies out of the total number of strategies to participate in the game. Each agent chooses the S strategy out of the total strategies. The minority game theory was shown to be computer simulation that the strategy with proper memory is more efficient than if it had a lot of strategies or actions. The traditional game model is a game in which the pay-off varies depending on the choice of cooperation (C) and defeat (D) [65]. If the game player can make two choices, the gain is different depending on the player's choice. When mutual consensus is reached, the gain is given R (reward) or P (punishment), and when the two men's choices are different, the gain of S (sucker) or T (temptation) is obtained [65]. Because Pareto is deficient, R> P is more advantageous than when two agents cooperate with each other. In a game of Stag Hunt (R> T> P> S), the situation is the problem but not greed. However, in the Chicken game (T> R> S> P), the problem is greed but not fear. In the well-known Prisoner's Dilemma game the payoff satisfies the relation. In an actual social phenomenon, an evolutionary game in which a large number of game participants choose an evolutionary strategy of a game has also been actively studied recently [65]. Similar to Prisoner's dilemma game, research on the coevolutionary public good game has been extensively conducted [66].

3.4 Social Flow Models

The heterogeneous interactions heavily influence on flocking dynamics [67]. Agent-based models have been actively applied to the phenomenon of crowding of humans, animals, and insects. Social flow models that simulate the behavior of agents in social phenomena have been extensively studied. When a fire occurs in a building, people are afraid to try to escape to the exit. When a suspected terrorist explosion occurs in a large venue or a sports arena, many people are trapped at the exit at once and often fail to escape or witness massive plague accidents. This phenomenon is called "the faster is slower effect" [68]. People and animals in fear do not behave rationally, but rather follow on their perceived surroundings and their behavior. Behavior following neighbors' behavior causes unexpected problems. The collective motion of agents is an object that can apply the complex system method to the movement out of equilibrium [69]. The heterogeneous interactions are heavily influenced by flocking dynamics [67].

The study of human flow in society can be divided into rule-based models and force-based models based on Cellular Automata. A typical example of a rule-based model is the Nagel-Schreckenberg model (NaSch model) that implements the traffic congestion phenomenon on the road with Cellular Automata [70]. This model has been extensively applied to study for understanding traffic congestion [71-73]. The Nasch model is a system of interacting particles driven far from equilibrium. In the NaSch model, the highway is treated as a one-dimensional grid consisting of cells. Each cell is filled with automobiles or is empty. Depending on the discrete time, the car is a cellular automata that progresses in four stages: acceleration, deceleration, random deceleration of the driver, and forward movement. This model explains the traffic congestion on the highway. In the NaSch model, the maximum speed of the vehicle is $v_{max}$ and the speed can have an integer number from 0 to $v_{max}$. If the maximum velocity is $v_{max} = 1$, then the NaSch model is like a far from equilibrium driven interactive system, so called, totally asymmetric simple exclusion process [74,75]. The NaSch model describes the phantom traffic jam where traffic congestion occurs despite the absence of a car accident when the vehicle density on the road increases. In 2000, Helbing *et al* simulated human behavior in the form of an agent-based model where many people come together at the exit of a building [76]. The evacuation model of Helbing et al is a representative forced-based model. It simulate movement by dynamically describing the interaction between people acting in the building [77,78]. Understanding the behavior of a person in an emergency can be a great help in real life because he can figure out strategies to escape without being blocked [79]. Helbing's escape model simulated the phenomenon that people could not get out of the way when they were driven together at one door. To solve this problem, one showed that it is possible to alleviate the clogging phenomenon when the column is in front of the door. Transportation and escape models can be easily extended to pedestrian models that simulate human movement on streets or in certain restricted spaces [77]. Ma *et al* show the dual effects of pedestrian density in building on the evacuation dynamics. The visibility and the exit limit in building have dual effects to evacuate [78]. Ha and Lykotrafitis consider agent-based modeling of a multi-room multi-floor building emergency evacuation [79]

3.5 Opinion Dynamics and Voter Models

The opinion dynamics is a field in which agent-based models exert a strong force. Opinion decisions are needed in various social contexts [80]. When agents decide their opinions, not only are each agent affected by other agents around, but also influenced by social media such as broadcasting, newspapers and social networking. Agents maintain their own current opinions and change their opinions by the influence of neighbors and the environment. Voting in elections is an example of representative opinion dynamics. In economic phenomena, people make a variety of choices, which can also be applied to opinion dynamics. For example, when people buy a computer, they can select the window as the operating system, or choose linux or Mac OS [1-5]. There are many cases in which we have to make a choice in social phenomena. The concept of statistical physics can easily be applied to the dynamics of opinion. If you have to choose one of two opinions, you can think of each agent as spin up or spin down. When treating the Ising model as a simple opinion dynamic model, agents are more strongly influenced by the majority state of the interacting agents [26,28-30].

The agent chooses a neighboring agent on the lattice randomly [81]. When agents have different opinion, they switch their opinions to ++ or -- status with a 1/2 chance. Thus, the order parameter of the system is the density of the + - pair $\rho(t)$. In spin dynamics, the magnetization is $m(t)$ and $\rho(t) = (1 - m(t))/2$. The state where all agents reach full consensus becomes + or - absorbing state of the system [26]. A variety of voter models can be considered depending on how the chosen agent interacts with his neighbors. Consider a system consisting of N agents in a two-dimensional lattice. Each agent expresses two opinions as $S_i = \pm 1$. The simplest voter model follows the majority rule. One of the agents placed on the two-dimensional lattice is arbitrarily selected. This agent changes his opinion in proportion to the number of opinions of neighboring agents. In Fig. 2, when three of his neighbors have the same opinion and one is different, in the next step the agent sticks to the current opinion with a probability of 3/4 and replaces the opinion with a 1/4 probability [26].

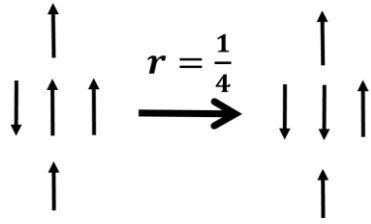

[Fig. 2] Majority voter model. The agent in the center is in the opinion up (left). At the update the agent sticks to the current opinion with a probability of r = 3/4 in the next step or changes his opinion to down with a probability of r = 1/4.

At the two-dimensional lattice in a position x, the transition probability $w_x(s)$ of the agent with opinion $s(x)$ is given by [26]

$$w_x(s) = \frac{1}{2}\left[1 - \frac{s(x)}{z}\sum_{y(x)} s(y)\right],$$

where $y(x)$ means the summation for the all nearest neighbors of $x$ and $z$ is the coordination number. The asymptotic behaviors of the order parameter in this voter model are given by [26]

$$\rho(t) \approx \begin{cases} t^{-1/2}, & d = 1 \\ (\ln t)^{-1}, & d = 2. \\ O(1), & d > 3 \end{cases}$$

Below two dimensions, the +- pair disappears after a long time and eventually a consensus is reached. The two dimension is the marginal dimension, and the order parameter is reduced to $(\ln t)^{-1}$, resulting in consensus. On the other hand, if the spatial dimension is more than three dimensions, the order parameter converges to a finite value. That is, the whole system does not converge on one side of the opinion. There are many opinions, but there are a certain percentage of agents with minor opinions. The time it takes for an entire system with N agents to reach consensus is called $T_N$. In this simple voter model, consensus time is given by [26]

$$T_N \approx \begin{cases} N^2, & d = 1 \\ N\ln N, & d = 2. \\ N, & d > 3 \end{cases}$$

The voter model with power-law distribution of inter-event interval showed slow consensus on a ring compared with the exponential distribution of inter-event interval [82]. A transition between an active phase to a fragmented phase is observed in a voter model on a directed adaptive network with fixed out-degree distribution [83]. In the presence of perfectly partisan voters (like zealots), a controlling strategy is proposed to maximize the share of a party in a social network of independent voters [84]. In the noisy voter model, the degree heterogeneity has a strong influence on the location of the critical point of a finite-size transition, on the local ordering of the system, and on the temporal correlation [85]. Jedrzejewski obtain a critical point of the q-voter model with independence on complex networks by using the pair approximation [86]. When the agent is navigating on the complex networks, the formation of consensus is possible for the nonlinear interactions between the agents [87].

4. Platforms of ABM

We need programming language and simulation platform to use when simulating an agent-based model. Commonly used languages for computer simulation programs are C-language, Fortran, Java, Python, MATLAB, and Mathematica. Choose the language you are most familiar with and program it. If you are using an agent-based model platform, you can consider Netlogo, Repast, Swarm, and Mason. [88-90]. When simulating an agent-based model that requires complex and large-scale computations, it is good idea to use the low level programming language. If the system is small and do not need fast calculations, you can utilize a variety of agent-based model platforms for simulation while real-time checking results graphically.

There are about 85 known agent-based model toolkits [89]. Netlogo (https://ccl.northwestern.edu/netlogo/) is an ABM toolkit that is easy to use and can be used as an educational tool. Netlogo consists of a menu-type platform and provides libraries with various examples [91]. The Netlogo built-in library provides a well-known ABM model that allows users to easily use the simulation platform. However, large-scale simulations and complex ABM simulations are difficult to implement. Mason (https://cs.gmu.edu/~eclab/projects/mason/) is easy to use for experienced researchers with programming experience, and it is easy to implement ABMs with large system sizes [88]. It is especially suited for simulations that require multi-agent ABM and long computations. However, since it is not a user-friendly toolkit, it is suitable for users who want fast calculation. In particular, it does not provide terminal windows for users, and there is no debugging tool. Repast (https://repast.github.io/) runs faster than other platforms and is a java platform [88]. Repast provides classes for various geographical and network functions. However, since Repast does not provide a user-friendly environment, it is suitable for users who are more programming-experts than first-time users. Choose an appropriate platform based on your ability to be a beginner or a professional programmer. When choosing a platform, it chooses a platform suitable for itself considering fagents such as execution speed, maximum size of system size that can be implemented, user-friendly environment, provision of graphical environment, and provision of debugging tools.

5. Conclusions

The agent-based models applied to social physics system were reviewed. We looked at the characteristics of ABM's components and agents. Agents have autonomy, behavioral rules, interactions with other agents, irrationality, and heterogeneity. Agents have memories of the past during the simulation and adapt through learning. We have looked at various examples of social complexity. ABM, which implements human behavior, can be classified into flow simulation, organizational simulation, economic market, and diffusional system. We examined the Sugarscape model and the artificial economic market in the ABM of econophysics. We examined the characteristics of minority game and majority game which are typical examples of game theory ABM. We examined crowding behavior, evacuation, traffic congestion, and pedestrian dynamics in social flow ABM. The ABM is widely applied to opinion dynamics and voter model. We introduced various platforms that can easily implement ABM and compared advantages and disadvantages. The ABM is applied to various fields such as social physics, economics physics, engineering, and ecological systems, and its utilization will increase.

Acknowledgements

This work was supported by INHA UNIVERSITY Research Grant.